\documentclass[11pt,a4paper]{article}
\usepackage[T1]{fontenc}
\usepackage{geometry}
\usepackage{amsmath}
\usepackage{graphicx}
\usepackage{amssymb}
\usepackage{esint}
\usepackage{xspace}
\usepackage{epsfig}

\newcommand{\lwig}{\mbox{\;\raisebox{.3ex}
    {$<$}$\!\!\!\!\!$\raisebox{-.9ex}{$\sim$}\;}}
\newcommand{\gwig}{\mbox{\;\raisebox{.3ex}
    {$>$}$\!\!\!\!\!$\raisebox{-.9ex}{$\sim$}}\;}

\newcommand{\CnuB}{C\(\nu\)B\xspace}
\def\be{\begin{equation}}
\def\ee{\end{equation}}
\def\bea{\begin{eqnarray}}
\def\eea{\end{eqnarray}}
\def\m{\m_{\nu}}

\makeatletter

\newcommand{\lyxdot}{.}

\makeatother

\begin{document}

\title{\vspace*{-1cm}{\hfill\normalsize \tt DESY 07-174}\vspace*{+1cm}\\
Relaxing neutrino mass bounds by a running cosmological constant}
\author{Florian Bauer and Lily Schrempp\\
\\
\textit{Deutsches Elektronen-Synchrotron DESY}\\
\textit{Hamburg, Notkestr. 85, 22607 Hamburg, Germany}\\
\\
Email: \texttt{florian.bauer@desy.de}, \texttt{lily.schrempp@desy.de}}
\date{5 November 2007}
%\address{$^1$ Deutsches Elektron-Synchroton DESY, Hamburg, Notkestr. 85, 22607 Hamburg, Germany}

%\eads{\mailto{florian.bauer@desy.de},
%\mailto{lily.schrempp@desy.de}}
%\date{{\today}}
\maketitle
\begin{abstract}
We establish an indirect link between relic neutrinos and the dark energy sector which originates from the vacuum energy contributions of the neutrino quantum fields. Via renormalization group effects they induce a running of the cosmological constant with time which dynamically influences the evolution of the cosmic neutrino background. We demonstrate that the resulting reduction of the relic neutrino abundance allows to largely evade current cosmological neutrino mass bounds and discuss how the scenario might be probed by the help of future large scale structure surveys and Planck data.
\end{abstract}

\maketitle

\section{Introduction}

Two of the outstanding advances of the last decades were the confirmation of neutrino oscillations and the observation of a late-time acceleration of the cosmological expansion attributed to a form of {\it dark energy}, its simplest origin being a Cosmological Constant (CC). In this work, we discuss the implications for cosmological neutrino mass bounds arising from an interaction between the CC and relic neutrinos.

According to Big Bang theory the Universe was merely about one second old, when cosmic neutrinos fell out of equilibrium due to the freeze-out of the weak interactions. From the standard presumption that the neutrinos have been non-interacting ever since decoupling, it follows that their distribution is frozen into a freely expanding Fermi-Dirac momentum distribution. Consequently, these neutrinos are assumed to homogenousely permeate the universe as the cosmic neutrino background (\CnuB) with a predicted average number density of $\sum n_{\nu_{0}}=\sum n_{\bar{\nu}_{0}}=168\,{\rm{cm}}^{-3}$ today. This substantial relic abundance only falls second to the photons of the cosmic microwave background (CMB), the exact analog of the \CnuB. However, owing to the feebleness of the weak interaction, all attempts to directly probe the \CnuB in a laboratory setting have so far been spoiled~\cite{Hagmann:1999kf,Ringwald:2004np,Gelmini:2004hg,Ringwald:2005zf}. Yet, evidence for its existence and knowledge of its properties can indirectly be inferred from cosmological measurements sensitive to its presence like the abundance of light elements as well as CMB and Large Scale Structure (LSS) measurements (see e.g.\ Ref.~\cite{Lesgourgues:2006nd} for a recent review). For example, LSS data can reflect kinematical signatures of relic neutrinos which arise, since on scales below their free streaming scale massive neutrinos cannot contribute to the gravitational clustering of matter. Through the metric source term in the perturbed Einstein equations, this lack of neutrino clustering translates into a small scale suppression of the matter power spectrum which depends on the fractional energy density provided by neutrinos to the total matter density. Hence, within the standard cosmological model, LSS data allows to infer bounds on the absolute neutrino mass scale $\sum m_{\nu}=0.2 \div 1$ eV (2$\sigma$) depending on the employed data sets (see e.g.~\cite{Goobar:2006xz,Seljak:2006bg,Feng:2006zj,Cirelli:2006kt,Hannestad:2006mi,Fogli:2006yq,Spergel:2006hy,Tegmark:2006az,Zunckel:2006mt,Kristiansen:2006ky}). However, these bounds are in tension with the claim of part of the Heidelberg-Moscow collaboration of a $>4\sigma$ evidence for neutrinoless double beta decay translated into a total neutrino mass of $\sum m_{\nu}>1.2$ eV at $95\%$ c.l.~\cite{KlapdorKleingrothaus:2004wj}. While this claim is still considered as controversial (see e.g.~\cite{Elliott:2004hr}), at present other independent laboratory experiment are not sensitive enough to verify it. However, considering that cosmological neutrino mass bounds strongly rely on theoretical assumptions on the \CnuB properties such an apparent tension might hint at new physics beyond the standard model. Namely, it might imply the presence of non-standard neutrinos interactions which reduce the relic neutrino abundance and thus relax cosmological neutrino mass bounds (see Refs.~\cite{Lesgourgues:2006nd,Beacom:2004yd} and references therein). Hence, the improvement of current mass limit from tritium $\beta$ decay experiments, $\sum m_{\nu}<6.6$ eV ($2\sigma$)~\cite{Weinheimer:2003fj,Lobashev:2001uu}, by the approved tritium experiment KATRIN with a projected sensitivity of $0.2$ eV~\cite{Osipowicz:2001sq,KATRIN}, offers the exciting possibility to probe such new neutrino interactions in the near future. What kind of new interactions could neutrinos have?

Recently, it has been argued that a natural origin for new neutrino interactions would be the sector responsible for dark energy, a smooth energy component, which is assumed to drive the observed accelerated expansion of the universe~\cite{Gu:2003er,Fardon:2003eh,Amendola:2007yx,Wetterich:2007kr}. A scenario has been proposed~\cite{Fardon:2003eh}, in which the homogenousely distributed relic neutrinos are promoted to a natural dark energy candidate, if they interact through a new scalar force. As a clear and testable signature of this so-called Mass Varying Neutrino (MaVaN) Scenario the new neutrino interaction gives rise to a variation of neutrino masses with time. The resulting phenomenological consequences have been explored by many authors, in particular the implications for cosmology and astrophysics have been considered in Refs.~\cite{Li:2004tq,Brookfield:2005td,Brookfield:2005bz,Afshordi:2005ym,Ringwald:2006ks,Schrempp:2006mk,Friedland:2007vv,Bjaelde:2007ki,Bean:2007ny}.

In this paper we propose an alternative approach for the realization of a dynamical influence of the dark sector on relic neutrinos (and vice versa), which is complementary to the avenue in the MaVaN Scenario in the following sense. While we also start from the requirement of energy-momentum conservation of the coupled two-component system, we, however, require the neutrino masses to be constant. As a direct consequence, the energy exchange with the dark sector demands a non-conservation of the number of neutrinos and accordingly its variation with time. Such a set-up naturally arises on the basis of Einstein's theory of general relativity in the appropriate framework provided by quantum field theory on curved space time~\cite{Birrell:1982ix}. Namely, this description requires the running of the vacuum (or zero point) energy or equivalently the CC as governed by the renormalization group~\cite{Elizalde:1993ew,Elizalde:1994av,Shapiro:1999zt, Shapiro:2000dz, Guberina:2002wt, Grande:2007wj}, which implies an energy exchange between the CC and matter fields~\cite{Horvat:2005ua, Carneiro:2006yv, SilvaeCosta:2007je, Bilic:2007gr}. Within a suitable renormalization scheme, we find the relic neutrinos to govern the running of the CC at late times, when they turn non-relativistic. We analyse the consequences of the energy-momentum exchange between the relic neutrinos and the CC both for the late-time dynamics of the running CC and for the phase-space distribution and the abundance of relic neutrinos. Interestingly, independent of the absolute neutrino mass scale realized in nature, in the redshift range accessible by current cosmological probes we find the effective equation of state of the running CC to be extremely close to $-1$ as suggested by recent data (e.g.~\cite{Astier:2005qq}). However, we demonstrate that within the considered scenario the effects of the \CnuB on cosmological measurements are drastically reduced. As a consequence, current neutrino mass limits arising from cosmology can be largely evaded. Furthermore, we discuss how the considered scenario can be tested in the near future by the help of cosmological and laboratory based measurements. 

The outline of the paper is as follows. In Sec.~\ref{sec:RGE} we provide the theoretical framework for the considered scenario, and in Sec.~\ref{sec:Evolution} we analyse in detail the time evolution of the neutrinos and the CC arising from their interaction. Sec.~\ref{sec:Pheno} deals with the phenomenological consequences of the new neutrino interaction, in particular, with the implications for the inference of neutrino mass bounds from current LSS and CMB data. In Sec.~\ref{sec:Probing} we provide an outlook how the considered scenario can be tested both by future cosmological probes and by experiments sensitive to Pauli-Blocking effects arising from the presence of the \CnuB. 

\section{Running vacuum energy\label{sec:RGE}}

The existence of zero-point energies is an inevitable consequence of quantum field theory in which every quantized field is contributing to the vacuum energy density~$\Lambda$. In the absence of notable gravitational interactions, however, the CC can be completely ignored as it has no influence on non-gravitational physics. In contrast to this, the cosmological evolution can be very sensitive to the CC~\cite{Padmanabhan:2007xy}, and the impact of quantum corrections should be taken into account. Unfortunately, the calculation of the vacuum contribution~$\Delta\Lambda$ of a quantum field of mass~$m$ leads to a divergent result, which can be expressed as a momentum integral over all zero modes:~$\Delta\Lambda\sim \int\mbox{d}^{3}p\,\sqrt{p^{2}+m^{2}}$. Na\"{\i}ve estimations with a high-energy cut-off, e.g.\ at the Planck scale~$M_{\rm Pl}$, are far from successful since they lead to huge discrepancies with the measured value~$\Lambda_0$ of the CC:
\[ \Delta\Lambda \sim  M^4_{{\rm Pl}} \approx 10^{123} \Lambda_0. \]
Nevertheless, quantum field theory on a curved spacetime~\cite{Birrell:1982ix} allows us to obtain at least some information from the divergent contributions. In this framework gravity is considered to be a completely classical theory well below the Planck scale, which provides the background for quantum fields and their excitations. By using well-known procedures, infinities like~$\Delta\Lambda$ arising in this set-up can be made finite and eventually lead to meaningful results by treating them in an analogous way as for example divergent quantum corrections of the fine-structure constant in quantum electrodynamics~(QED). Consequently, regularization and renormalization techniques allow to formulate a renormalization group equation~(RGE) for the CC. Of course, this procedure does not determine the absolute value of~$\Lambda$, but only its change with respect to an initial value~$\Lambda_{0}$ that has to be measured by experiments. Similar to the running fine-structure constant, the vacuum energy~$\Lambda(\mu)$ becomes a scale-dependent variable with~$\mu$ as a characteristic energy scale, called the renormalization scale.\footnote{One also finds a RGE for Newton's constant~$G$, but its running
is suppressed by the ratio~$(m/M_{{\rm Pl}})^{2}$. Therefore
we can safely ignore this scale-dependence in this work.} For each bosonic/fermionic field degree of freedom~(DOF) the RGE can be expressed in the form
\begin{equation}
\mu\frac{\partial\Lambda}{\partial\mu}=\pm\frac{m^{4}}{32\pi^{2}},\label{eq:QF-RGE}
\end{equation}
where~$m$ denotes the mass of the corresponding quantum field. The complete equation is then given by the sum of DOFs of all fields in the theory. For free fields Eq.~(\ref{eq:QF-RGE}) is an exact result, and even in curved space-time there are no higher order corrections. Unlike QED, where the renormalization scale~$\mu$ is usually related to external momenta or energy scales, the running of~$\Lambda(\mu)$ originates from a zero-point function
and does not involve external scales. Therefore, we have to find a suitable identification for~$\mu$. In cosmology several choices have been discussed before in the literature, e.g.\ the Hubble expansion rate~$H$ or quantities related to horizons, see \cite{Babic:2004ev,Bauer:2005rpa,Bauer:2005tu} and the references therein. However, in contrast to such global choices for $\mu$, in this work we investigate a scale that is more characteristic for the contributing field and the corresponding particle spectrum. 
To motive its choice, it is important to note that Eq.~(\ref{eq:QF-RGE}) has been derived in the~$\overline{{\rm MS}}$ renormalization scheme that usually does not show decoupling effects at energies sufficiently below the mass~$m$~\cite{Babic:2001vv,Gorbar:2002pw,Shapiro:2003ui}. Since the current CC is much smaller than $m^4$ for most particle masses, the na\"{\i}ve evolution of the RG running down to low energies~$\mu\ll m$ might induce unrealistically huge changes in the CC at late times. Instead of stopping the RG evolution by hand in the low energy regime, we choose our RG scale such that it naturally implements some sort of decoupling when~$\mu$ approaches the field mass~$m$.

Before identifying the renormalization scale, we have to note first that a time-dependent~$\mu$ generally implies that the vacuum energy~$\Lambda$ cannot remain constant anymore, even though its fundamental equation of state is still~$-1$. Therefore, in order to have a consistent framework, Bianchi's identity has to be satisfied. On a cosmological background this means
\begin{equation}
\dot{\Lambda}+\dot{\rho}+3H(\rho+P)=0,\label{eq:QF-Bianchi}
\end{equation}
where the dot denotes the derivative with respect to cosmic time~$t$. For concreteness, let~$\rho$ and~$P$ be the total energy density and pressure of all components apart from the CC. Since for self-conserved matter the term $\dot{\rho}+3H(\rho+P)$ vanishes, there must obviously exist a non-trivial interaction between~$\Lambda$ and~$\rho$ accounting for the exchange of energy-momentum between both sectors. Consequently, both the matter energy density and particle number density do not obey the standard dilution laws
anymore. Note that this is qualitatively different from self-conserved dark energy sources, like (non-interacting) scalar field or holographic dark energy models.

Now, one might ask which part of the total matter energy density consisting
mainly of cold dark matter, baryons, leptons and photons is actually exchanging energy-momentum with the running CC to compensate for the time dependence of $\Lambda$ in Eq.~(\ref{eq:QF-Bianchi}). Since the microscopic realization of the implied interaction is not known, let us in the following motivate our approach to answer this question. The vacuum contribution of a quantum field of mass $m$ leads to a RGE as specified in Eq.~(\ref{eq:QF-RGE}). At the same time, the quantum field describes particles of mass $m$, whose energy density contributes to the total matter density $\rho$. Accordingly, a natural way of satisfying Eq.~(\ref{eq:QF-Bianchi}) seems to be provided if the system of the vacuum contribution and the energy-momentum of the quanta arising from the \emph{same} quantum field is conserved. Correspondingly, each field obeys an equation of the form of Eq.~(\ref{eq:QF-Bianchi}), however, with~$\Lambda$, $\rho$ and $P$ belonging to this field only. Following the same line of reasoning, the RG scale~$\mu$ should be related to this self-conserved system, too.

To examine the RG evolution of a weakly interacting quantum field, in our framework it is appropriate to introduce the momentum distribution function~$f$, which characterizes the density of the corresponding particles in a given momentum bin on the homogeneous background of the universe. By its help, the energy density, pressure and the particle number density of each field DOF can respectively be expressed as
\begin{eqnarray}
\rho & = & \frac{T^{4}}{(2\pi)^{3}}\int\mbox{d}^{3}y\,\omega(y)f(y),\,\,\,\,\,\omega(y):=\sqrt{y^{2}+\frac{m^{2}}{T^{2}}},\nonumber \\
P & = & \frac{T^{4}}{(2\pi)^{3}}\int\mbox{d}^{3}y\,\frac{y^{2}}{3\omega(y)}f(y),\label{eq:QF-rPn}\\
n & = & \frac{T^{3}}{(2\pi)^{3}}\int\mbox{d}^{3}y\, f(y).\nonumber
\end{eqnarray}
Here, we take~$f$ to depend on~$y:=p/T$, the ratio of the momentum~$p$ and the effective temperature~$T$ characterizing the species, which in the considered framework have the same scaling with redshift such that~$y$ is constant in time. Hence, while the effective temperature $T$ is cooled by the expansion, the form of such a distribution is maintained in the absence of interactions. However, as it will turn out later on, taking into account the interaction with the CC, the distribution function will not expand freely anymore but also change with time. 

Accordingly, a reasonable choice of the RG scale seems to be the average energy~$\rho/n$ of the system, since it behaves at high energies like the characteristic momentum or temperature scale $T$. Furthermore, it is bounded from below by the field mass~$m$, thereby implying the end of the RG running. In order to keep the spin statistics intact, we also implement the factors~$(1\pm f)$ for bosonic/fermionic fields,
\begin{equation}
\mu=\frac{\hat{\rho}}{\hat{n}},\,\,\,\,\hat{\rho}:=\frac{T^{4}}{(2\pi)^{3}}\int\mbox{d}^{3}y\,\omega f(1\pm f),\,\,\,\hat{n}:=\frac{T^{3}}{(2\pi)^{3}}\int\mbox{d}^{3}y\, f(1\pm f).
\label{eq:QF-mu}
\end{equation}
Note that by these means~$\hat{\rho}$ represents the average energy-density of the system which is available for the interaction with the CC. Furthermore, we would like to point out that for more general systems or sets of individual particles, the resulting form of Bianchi's identity and the RG scale might look very different from the considered case.

\section{The RG evolution of $\Lambda$ and the C$\nu$B\label{sec:Evolution}}

In the following, we restrict our discussion to homogeneously distributed matter in the universe, that is to CMB photons, to background neutrinos and in principle to possibly existing other relics. The photons in the CMB are massless and thus not affected by RG effects. However, relic neutrinos, should be subject to RG running since they are massive and very weakly interacting. According to Big Bang theory, their distribution was frozen into a freely expanding, ultra-relativistic Fermi-Dirac (FD) form,~$f_{0}(y)=1/(\exp(y)+1)$, when neutrinos decoupled from the thermal bath at a temperature scale of $T_{{\rm dec}}\sim1\,\mbox{MeV}\gg m$. In the following, we will therefore use this form as initial condition for~$f$ and calculate its time evolution by virtue of Bianchi's identity~(\ref{eq:QF-Bianchi}) and the integrated form of the RGE~(\ref{eq:QF-RGE})
\begin{equation}\label{IntRGE}
\Lambda(\mu)=\Lambda_{0}-\frac{m^{4}}{32\pi^{2}}\ln\frac{\mu}{\mu_{0}}.
\end{equation}
Let us stress that the following considerations generically hold for any weakly interacting fermionic species of mass~$m$ exchanging energy-momentum with the CC, if its initial distribution function assumes the ultra-relativistic FD form~$f_{0}(y)$. For this reason, and for simplicity, we will omit the index $\nu$ in the following, but reinstate it when we investigate the phenomenology of interacting neutrinos in Sec.~\ref{sec:Pheno}. Accordingly, without loss of generality, we have taken both~$\rho$,~$n$ and~$P$ in Eqs.~(\ref{eq:QF-rPn}) as well as $\Lambda(\mu)$ to correspond to one fermionic DOF, since the actual number of fermionic DOF drops out in Eq.~(\ref{eq:QF-Bianchi}). Here, the RG running of the mass can be neglected due to the absence of substantial interactions with other fields.

Because of the momentum integrals in Eqs.~(\ref{eq:QF-rPn}), the Bianchi identity~(\ref{eq:QF-Bianchi}) yields a rather complicated non-linear integral equation for the time-evolution of~$f(y)$:
\begin{eqnarray*}
0=\dot{\rho}+\dot{\Lambda}+3H(\rho+P) & = & \frac{T^{4}}{(2\pi)^{3}}\int\mbox{d}^{3}y\,\omega\Big[\dot{f}\\
 & - & \frac{m^{4}}{32\pi^{2}}\dot{f}(1-2f)\frac{1}{\hat{\rho}}\left(1-\frac{\hat{\rho}}{T\omega\hat{n}}\right)\\
 & + & \frac{m^{4}}{32\pi^{2}}f(1-f)\frac{3H}{\hat{\rho}}\left(\frac{y^{2}}{3\omega^{2}}+1-\frac{\hat{\rho}}{T\omega\hat{n}}\right)\Big].
\end{eqnarray*}
While the interaction tends to deform $f$ away from its equilibrium form~$f_0$ as discussed in the following, we will still use the effective temperature $T=T_{0}(z+1)$ as a measure for cosmic redshift as in the standard case. Accordingly, the present characteristic temperature of FD distributed neutrinos is~$T_{0}\approx 1.7\cdot10^{-4}\,{\rm eV}$. To find a solution of the integral equation, we require the expression under the integral to vanish for any momentum value~$y$. This removes the outermost integral, and we obtain an equation that can be solved numerically. In terms of the derivative~$f'$ with respect to the redshift\footnote{$\dot{f}$ and~$f'$ are related by $\dot{f}(y)=-(z+1)Hf'(y)$.}~$z$ we find the following evolution equation
\begin{eqnarray}
f'(y) & = & \frac{3}{z+1}f(1-f)\cdot\frac{N}{D}\label{eq:QF-fprime},\,\,\mbox{where}\\
N & := & \frac{y^{2}}{3\omega^{2}}+1-\frac{\hat{\rho}}{T\omega\hat{n}},\\
D & := & \frac{\hat{\rho}}{(\frac{m^{4}}{32\pi^{2}})} +(1-2f)\left(\frac{\hat{\rho}}{T\omega\hat{n}}-1\right).\label{eq:QF-denom}
\end{eqnarray}
Let us in the following analyse this equation and compare it with our numerical results for the evolution of $f$. First, as a consequence of employing~$\hat{\rho}$ instead of~$\rho$ in Eq.~(\ref{eq:QF-mu}), one can see that the factor~$f(1-f)$ on the right-hand side ensures Pauli's principle to hold, $0\le f\le1$. Moreover, at early times, when~$\hat{\rho}\sim T^{4}\gg m^{4}$, the dynamics is strongly suppressed,~$|f'|\ll1$, and the neutrinos just behave as in the uncoupled case. During this time the denominator~$D$ is positive for any value of~$y$, while both the sign of the nominator~$N$ and of~$f'$ are $y$-dependent. For large momenta,~$y\rightarrow\infty$, we find~$f',N>0$ implying the decay of neutrinos. Considering that~$\mu$ is decreasing, this is the expected behavior following from Eq.~(\ref{eq:QF-Bianchi}). However, in the small momentum regime,~$y\rightarrow 0$, the distribution function increases because of~$f',N<0$. Since we are lacking a microscopic description of the dynamics of the coupled system, just from solving Bianchi's identity it is impossible to say whether the growing~$f$ owes to slowing down high-$y$ neutrinos or is also due to particle production.

\begin{figure}
\includegraphics[bb=99bp 270bp 306bp 674bp,clip,width=0.235\textwidth]{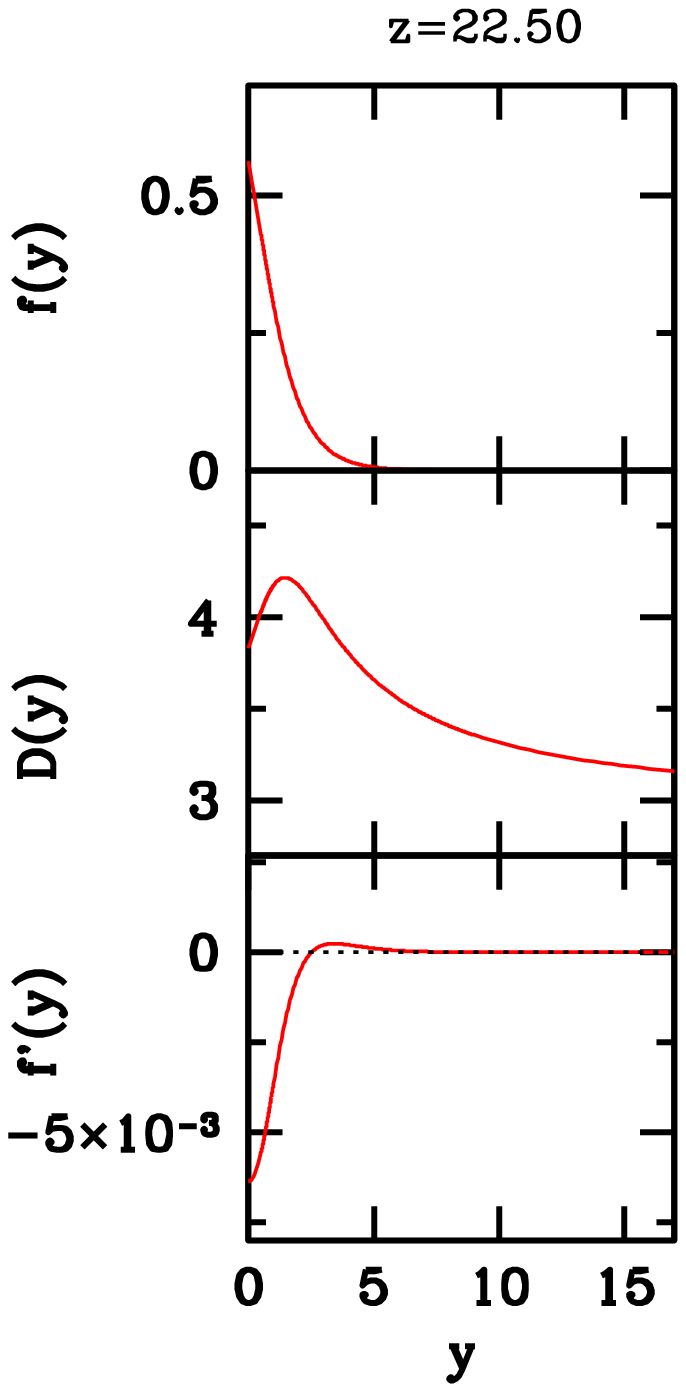}
\hfill{}
\includegraphics[bb=99bp 270bp 306bp 674bp,clip,width=0.235\textwidth]{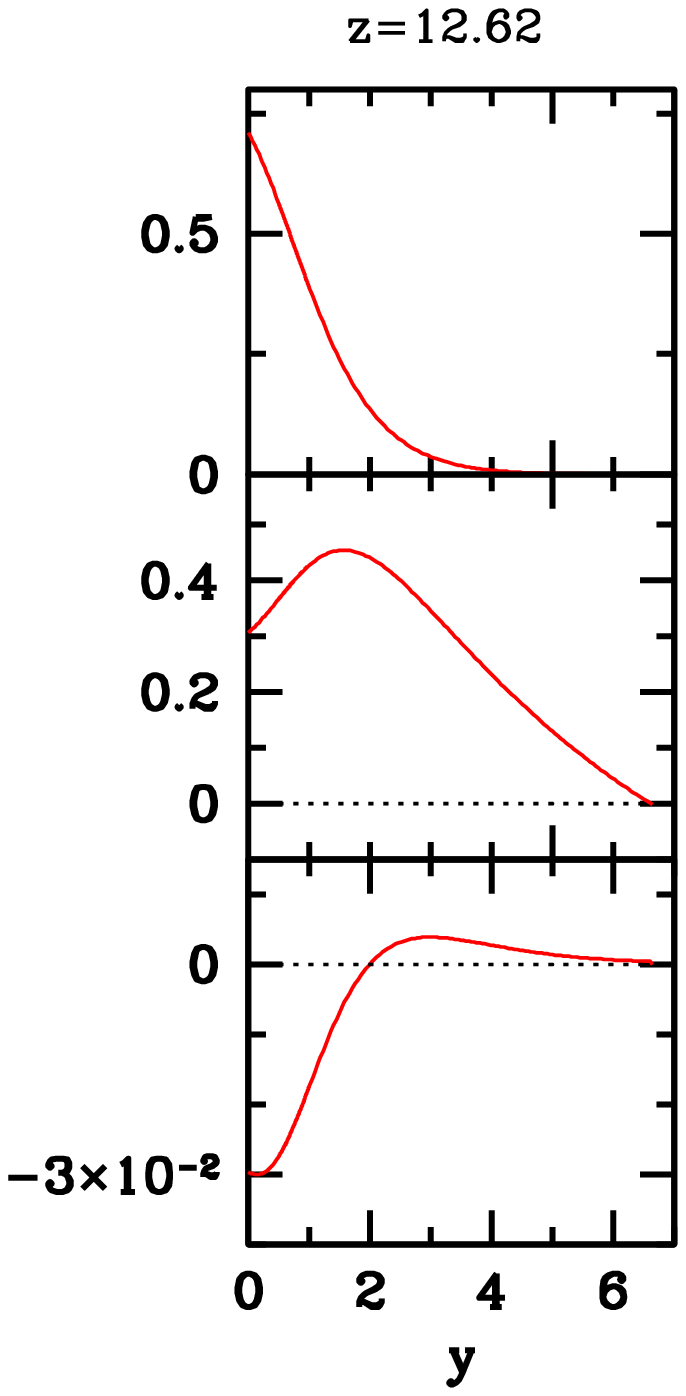}
\hfill{}
\includegraphics[bb=99bp 270bp 306bp 674bp,clip,width=0.235\textwidth]{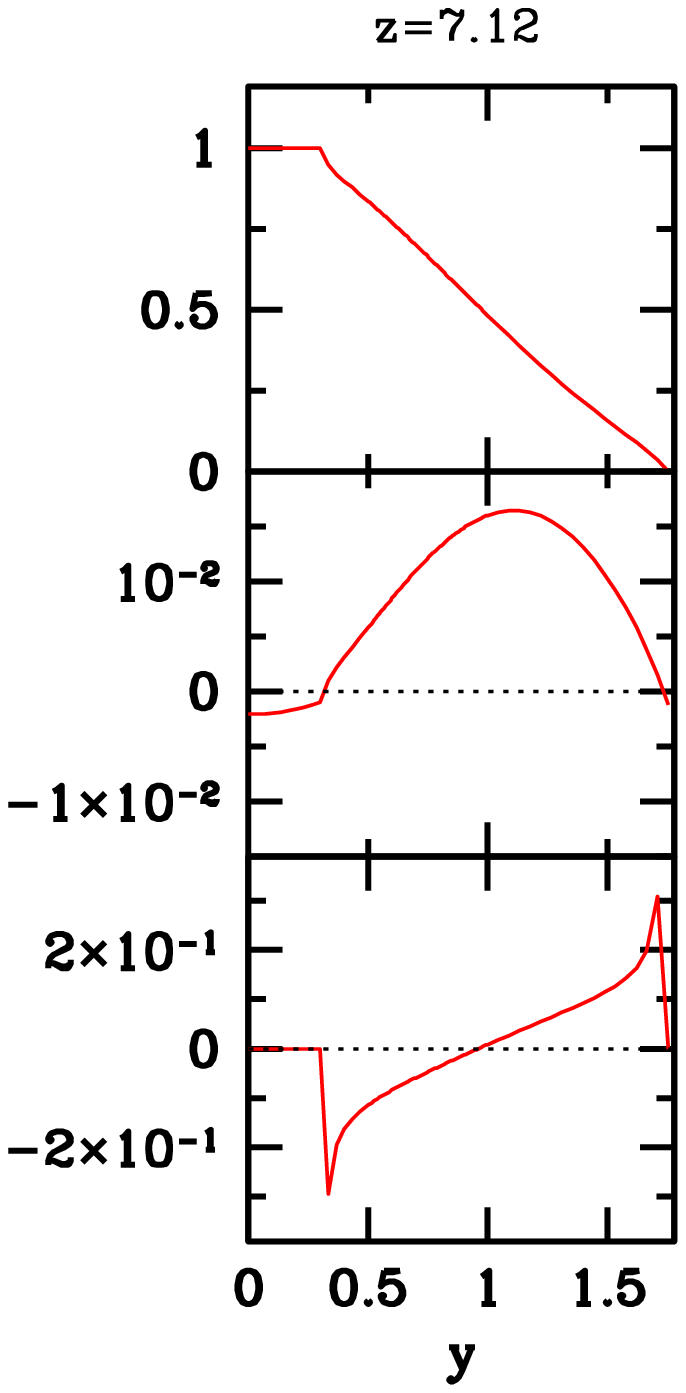}
\hfill{}
\includegraphics[bb=99bp 270bp 306bp 674bp,clip,width=0.235\textwidth]{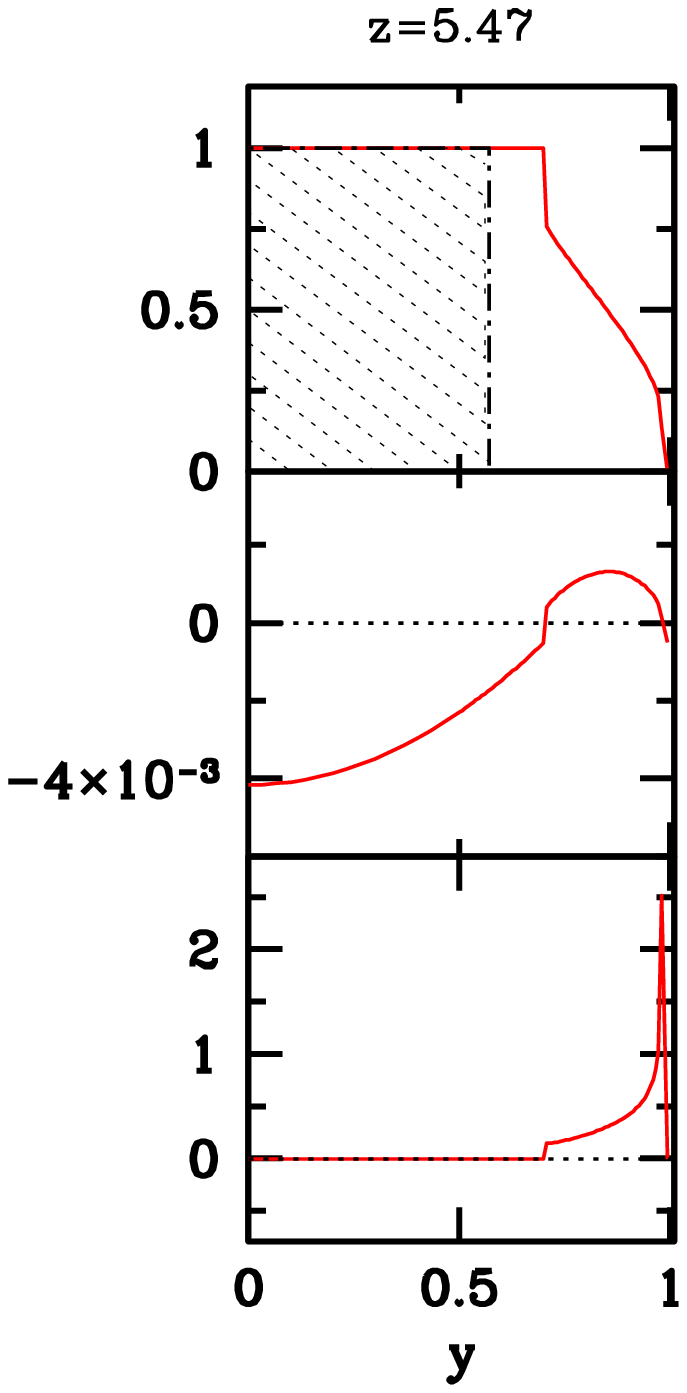}
\caption{\label{fig:QF-spectrum}Evolution of the momentum distribution profile~$f(y)$ of one neutrino with two DOFs and mass~$m_\nu=9\times 10^{-3}\,\rm{eV}$. Note the different scaling of the axis of ordinates and of the abscissae at each redshift~$z$. The plot corresponding to~$z=7.12$ demonstrates that the denominator $D(y)$ in Eq.~(\ref{eq:QF-denom}) exhibits two zeros, at which~$f(y)$ becomes~$1$ or~$0$, respectively. The final profile is acquired at~$z\approx4.75$. As indicated by the shaded region in the last~$z=5.47$ plot, it corresponds to a maximally degenerate Fermi sphere with all states occupied up to the Fermi momentum~$y_e\approx0.57$ (dashed-dotted line).}
\end{figure}

Finally, the distribution function is either approaching~$f=0$ or~$f=1$, when the positive denominator in Eq.~(\ref{eq:QF-fprime}) becomes zero. More precisely, the first zero appears for large momentum, where~$f(y\rightarrow\infty)\rightarrow0$ and~$\omega\rightarrow\infty$. Since the second term in Eq.~(\ref{eq:QF-denom}) in this limit is~$-1$, the time when this happens is determined by~$\hat{\rho}=\frac{m^{4}}{32\pi^{2}}$. This roughly corresponds to~$T\approx m/3$, i.e.\ the time when neutrinos turn non-relativistic. To arrive at this estimate, we have assumed that~$f$ has not changed much yet from its initial FD form $f_{0}$ and in addition that~$\hat{\rho}\approx\rho$. At the momentum position of this zero the spectrum is simply cut off with all larger momenta being removed, which is illustrated nicely in the last three plots in Fig.~\ref{fig:QF-spectrum} for decreasing redshift. 
This figure also demonstrates that for increasing time and further decreasing~$\hat{\rho}$, the momentum cut-off is moving to lower $y$-values.

A second zero in the denominator~$D$ will eventually occur for small values of~$y$. This happens when~$\hat{\rho}$ has decreased sufficiently and~$(1-2f)$ is negative. In the corresponding~$y$-range we find~$f',N<0$, and accordingly in this case~$f$ acquires its maximal value~$1$. As illustrated in the last of the two plots in Fig.~\ref{fig:QF-spectrum}, with decreasing redshift, the second zero of~$D$ occurs for larger and larger~$y$-values for which consequently~$f(y)\rightarrow 1$. Finally, both zeros in~$D$ meet at~$y=y_{ei}\approx 0.7$, which implies that $f$ cannot be evolved anymore according to Eq.~(\ref{eq:QF-fprime}).

At this stage the spectrum represents a maximally degenerate Fermi-Dirac distribution with all states completely occupied up to the Fermi momentum~$y_{ei}$. Once this Fermi sphere has formed, it is still possible to pursue the further time evolution of~$f(y)$ analytically by allowing the Fermi momentum~$y_{e}(t)$ to be time-dependent. As a result, we find respectively for the RG scale and the neutrino energy density
\begin{equation}
\mu=\frac{\hat{\rho}}{\hat{n}}\rightarrow\sqrt{y_{e}^{2}(t)T^{2}+m^{2}},
\,\,\,\rho=\frac{T^{4}}{2\pi^{2}}\int_{0}^{y_{e}(t)}{\rm d}y\, y^{2}\omega. \label{eq:QF-sph-mu-rho}
\end{equation}
Accordingly, other $f(y)$-dependent quantities can be determined by replacing the distribution function by a step function which is~$1$ for~$y\leq y_{e}(t)$ and~$0$ elsewhere. From Bianchi's identity we therefore obtain a much simpler differential equation for~$y_{e}(t)$,
\begin{equation}
\dot{y}_{e}(t)=\frac{Hy_{e}(t)}{1-16y_{e}(t)\frac{T^{4}}{m^{4}}\omega^{3}}\stackrel{\frac{m}{T}\gg y}{=}\frac{Hy_{e}(t)}{1-16\frac{T}{m}y_{e}(t)}.\label{eq:QF-ye-DGL}
\end{equation}
As can be read-off from Eq.~(\ref{eq:QF-ye-DGL}), since neutrinos are already non-relativistic and thus~$\frac{m}{T}\gg y$, $y_{e}(t)$ can only be evolved until it takes the final value~$y_{e}(t)\approx m/(16T)$. In our system this means that neutrinos are not able to exchange energy-momentum with the CC anymore. This is a direct consequence of including the Pauli principle in the RG scale, which blocks any further change in the coupled system. Therefore, it can be understood as the decoupling of the neutrinos from the running CC. Thus, henceforth, both the running of $\mu$ and of the CC is taken to cease such that the neutrinos evolve independently.

To find a (semi-)analytical expression for the final Fermi momentum~$y_e$, in the following, we will estimate the redshift~$T_{e}=T_{0}(z_{e}+1)$ corresponding to the decoupling time. To this end, we proceed by approximating the redshift evolution of interacting neutrinos. Accordingly, when the interaction effectively sets in, the temperature roughly satisfies~$T_{a}=m/3$, and we approximate the distribution function~$f(y)$ by a Fermi sphere with initial Fermi momentum~$y_{a}>y_{e}$. At this time,~$y_a$ can be determined by comparing the energy density~$\rho_{{\rm sph}}$ of this sphere with the energy density~$\rho_{{\rm FD}}$ of a standard FD distribution. In the non-relativistic neutrino regime, we obtain from the equality of
\begin{equation}
\rho_{{\rm sph}}\approx\frac{1}{6\pi^{2}}T_{a}^{3}my_{a}^{3}
\,\,\,\,{\rm and}\,\,\,\,
\rho_{{\rm FD}}\approx\frac{3\zeta(3)}{4\pi^{2}}T_{a}^{3}m,
\label{eq:QF-rho-sph-FD}
\end{equation}
the initial Fermi momentum~$y_{a}$ which takes the form~$y_{a} =(\frac{9}{2}\zeta(3))^{1/3}\approx 1.76$. By integrating the differential equation~(\ref{eq:QF-ye-DGL}) in the form\footnote{The general solution of Eq.~(\ref{eq:QF-yez-DGL}) for arbitrary~$y_{a,e}$ and~$T_{a,e}$ reads
\[
y_{e}=y_{a}\exp\left(\frac{m}{16}\left(\frac{1}{T_{a}y_{a}}-\frac{1}{T_{e}y_{e}}\right)\right).\]}
\begin{equation}
 y_{e}'(z)=-\frac{1}{z+1}\cdot\frac{y_{e}(z)}{1-16\frac{T}{m}y_{e}(z)}
\label{eq:QF-yez-DGL}
\end{equation}
from~$T_{a}=m/3$ to~$T_{e}$ with~$y_{e}=m/(16T_{e})$, we find~$y_{e}=y_{a}\exp(m/(16y_{a}T_{a})-1)\approx0.72$ and thus~$T_{e}\approx m/11.5$. Accordingly, the estimate for $T_{e}$ is of the same order of magnitude as the numerical result~$T_{e}\approx m/9$. Note that the final value,
\be\label{ye}
y_{e}=m/(16T_{e})\approx\frac{9}{16},
\ee
turns out to be independent of the mass parameter, which is confirmed by our numerical calculation.

Because the maximal physical momentum~$p_{e}=y_{e}T_{e}=\frac{m}{16}$ of the neutrino spectrum is much smaller than the mass~$m$, after decoupling from the CC, the neutrinos redshift just like non-relativistic dust, thus contributing to the dark matter. However, since they have transferred a substantial amount of energy-momentum to the vacuum energy during the interacting phase, with respect to non-interacting neutrinos of equal mass and energy density $\rho_{\rm FD}$, the magnitude of their energy density is reduced by a factor,
\begin{equation}
\frac{\rho_{{\rm FD}}}{\rho} = \frac{m}{m_{{\rm app}}} = \frac{\int_{0}^{\infty} {\rm d}y\,y^2 \frac{m}{T}/(e^y+1)}{\int_0^{y_e}{\rm d}y\,y^2\frac{m}{T}} \approx 30\,\,\,\,\mbox{for}\,\,T\leq T_e.\label{eq:QF-mapp}
\end{equation}
In the considered scenario, this relation also defines the apparent neutrino mass~$m_{{\rm app}}$, which follows from assuming the standard scaling law for the number density~$n_{{\rm FD}}$ and equal energy density, $\rho\simeq m_{{\rm app}}n_{{\rm FD}}$ for $T\leq T_e$.

\begin{figure}
\noindent \begin{centering}
\includegraphics[clip,width=0.5\textwidth]{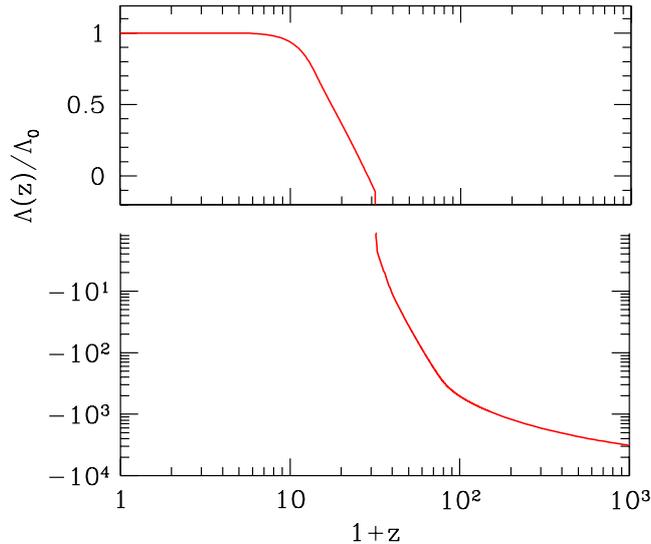}
\par\end{centering}
\caption{\label{fig:QF-LambdaVz}Vacuum energy density~$\Lambda(z)/\Lambda_{0}$ as a function of redshift for a hierarchical neutrino mass spectrum with $m_1=0$~eV, $m_2=9\times 10^{-3}$~eV, $m_3=5\times 10^{-2}$~eV and $2$ DOFs each. Note the different scaling of the ordinate.}
\end{figure}

Let us in the following qualitatively discuss the impact of the coupling between neutrinos and the CC on the background evolution. From the previous discussion we first note that the main effects of the interaction occur during an epoch set by the neutrino mass,
\begin{equation}
T_{e}\approx\frac{m}{9} < T < T_{a}\approx\frac{m}{3}.\label{QF-TaTe}
\end{equation}
Accordingly, the smaller the neutrino mass the later does the energy-momentum exchange between the coupled system occur. However, also the impact on the background evolution decreases with the neutrino mass, while the running of~$\Lambda$ becomes entirely negligible for~$m\ll\Lambda_{0}^{(1/4)}$ according to Eq.~(\ref{eq:QF-RGE}). In addition, in comparison to the~$\Lambda$CDM model with massless neutrinos, the energy density in light massive neutrinos is not much larger. As an example, we have plotted the redshift evolution of the cosmological constant in Fig.~\ref{fig:QF-LambdaVz} for a hierarchical spectrum with one massless and two massive neutrinos as allowed by neutrino oscillation experiments.
At high~$z$ both of the massive neutrinos contribute to the running of~$\Lambda$, while the heaviest neutrino of mass~$m_{3}=5\times 10^{-2}\,{\rm eV}$ is the dominant one. At~$z\approx30.6$ its momentum distribution has acquired its final, maximally degenerate form and thus the neutrino decouples. Afterwards, the second lightest neutrino with mass~$m_{2}=9\times 10^{-3}\,{\rm eV}$ is left over to drive the CC to its current value~$\Lambda_{0}$, which is reached at~$z\approx 4.75$. Neutrinos lighter than~$T_0 /9\approx 1.5\times 10^{-3}\,{\rm eV}$ could be interacting with the CC even today, but their influence on the running~$\Lambda$ would be too small to be detected.

However, we encounter a more pronounced effect of the interaction for larger masses, where~$\Lambda$ was very negative in the past and evolves to its current observed value~$\Lambda_{0}>0$. A negative CC might first look strange, but there are many potential sources for negative vacuum energy, e.g.\ spontaneous symmetry breaking~\cite{Shapiro:1999zt} or supersymmetry~\cite{Weinberg:1982id}. In addition, it turns out to be harmless since in the relativistic regime, where~$T\gg m$ and $\mu\simeq T$ the vacuum contribution is given by
\[
\Lambda_{0}-\Lambda=\frac{m^{4}}{32\pi^{2}}\ln\frac{\mu}{\mu_{0}}=\frac{m^4}{32\pi^{2}}\ln\frac{T}{m}.
\]
Obviously, this is much smaller than the energy density~$\simeq T^{4}$ of the dominant radiation components at early times. Therefore, the decay of the neutrinos and the influence of the running CC on the background evolution is negligible during this stage, thus leaving Big Bang Nucleosynthesis (BBN) unaffected by this mechanism. Only in the redshift range, where the energy exchange with the neutrinos is relevant,~$\Lambda$ may have a larger impact on the background evolution. To illustrate the effect we consider the case of three degenerated neutrinos with mass~$m_\nu=2.2\,{\rm eV}$ and $2$ DOFs each, as maximally allowed by $\beta$-decay experiments.

\begin{figure}[t]
\noindent \begin{centering}
\includegraphics[bb=56bp 139bp 550bp 718bp,clip,width=0.525\textwidth]{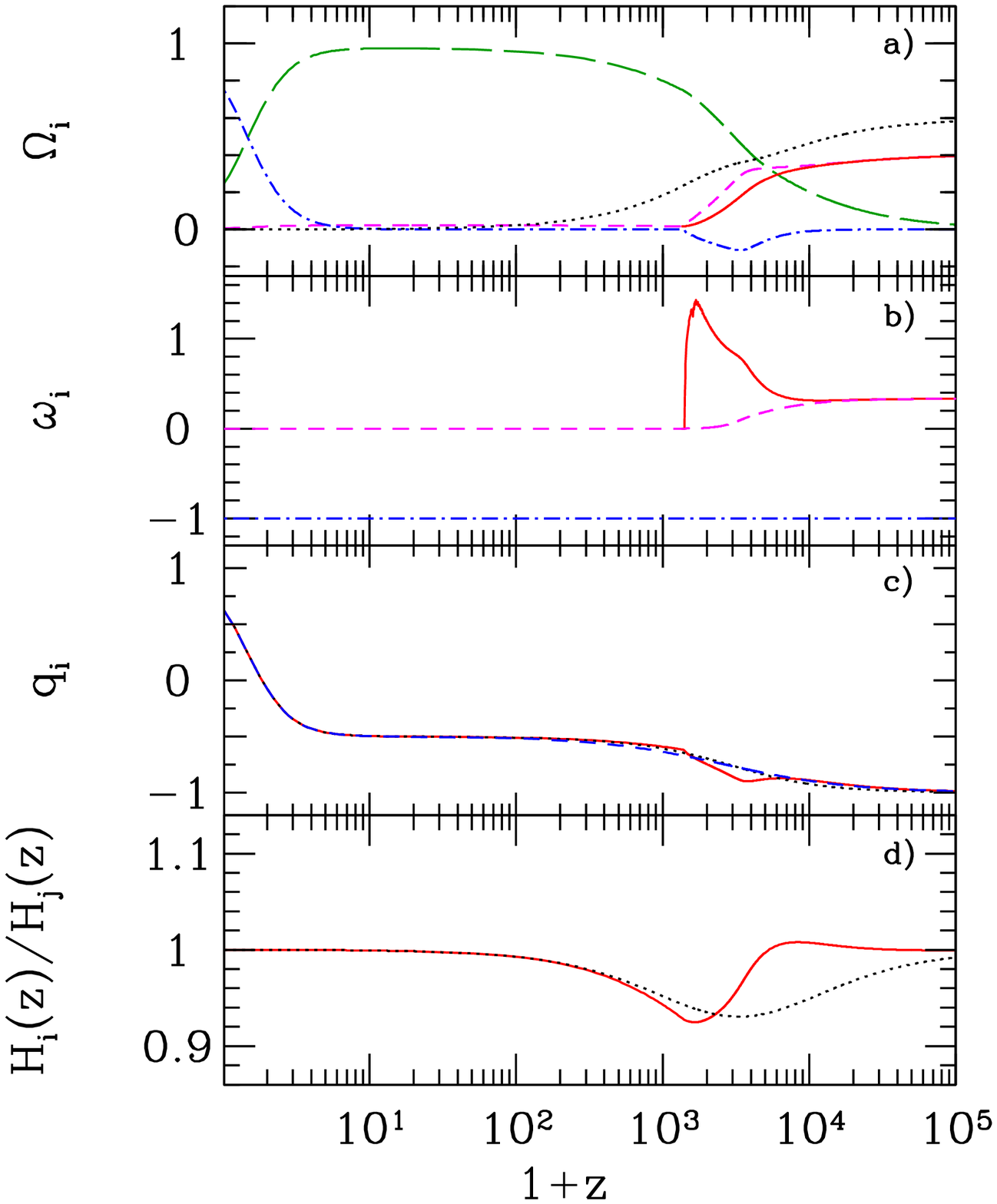}\hfill{}\includegraphics[bb=112bp 139bp 550bp 718bp,clip,width=0.465\textwidth]{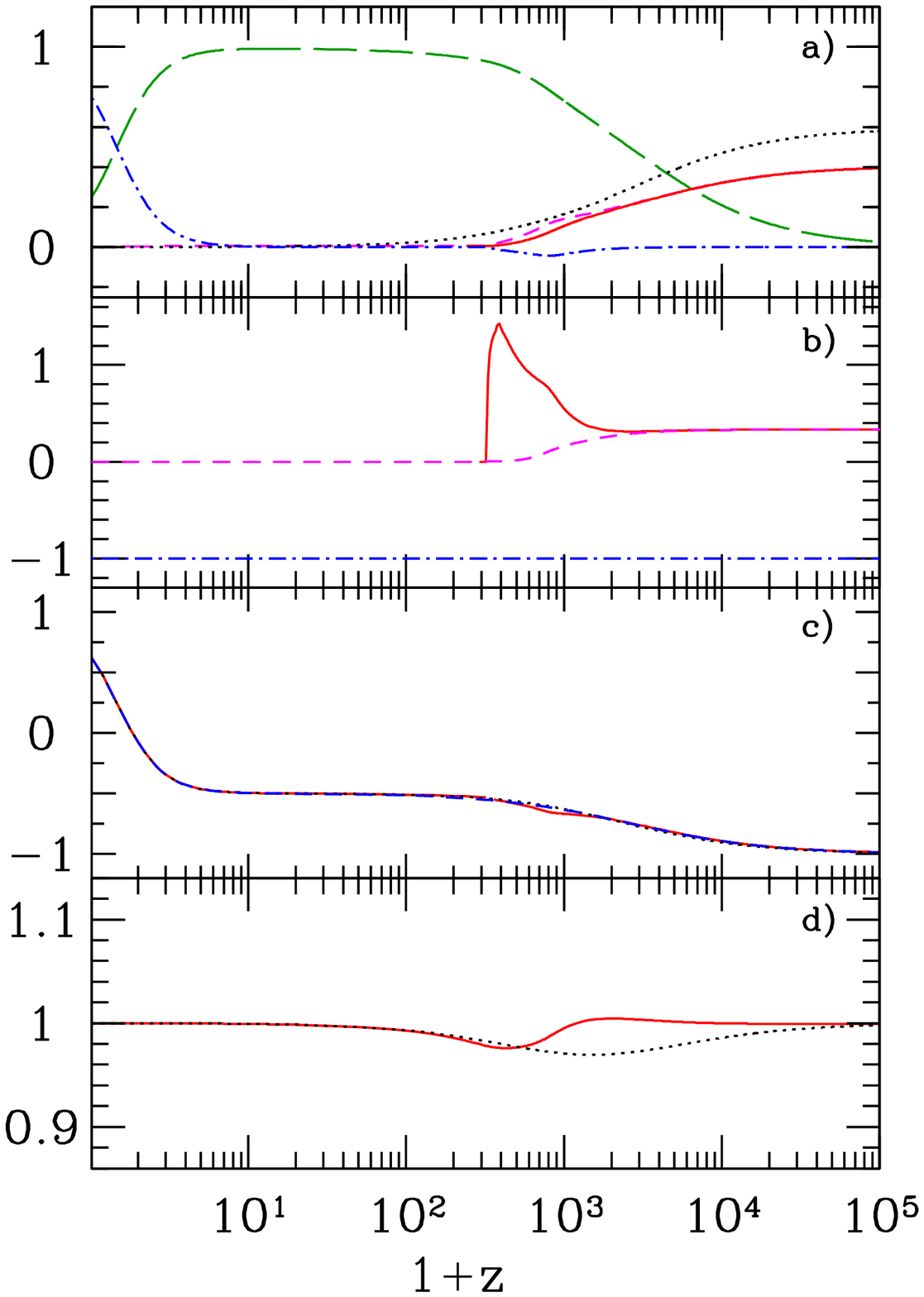}
\par\end{centering}
\caption{\label{fig:QF-OmegaVz} Cosmological evolution in the case of three degenerate neutrinos with two DOFs and mass~$m_\nu=2.2\,\rm{eV}$ (left) and~$m_\nu=0.5\,\rm{eV}$ (right) each. $a)$ The relative energy densities~$\Omega_{i}$ of the combined neutrino-CC system (solid), neutrinos (short-dashed), the CC (dashed-dotted), matter (long-dashed), radiation (dotted), respectively. $b)$ The effective EOS~$\omega_{\rm{eff}}$ of the combined neutrinos-CC system (solid) in the interacting phase, the EOS of neutrinos (dashed) and the CC (dashed-dotted), respectively. c) The acceleration factor~$q=\frac{\ddot{a}a}{\dot{a}^{2}}$ within $\Lambda$CDM (dotted), $\Lambda$MDM (dashed) and $\Lambda$MDM with neutrino-CC interaction (solid), respectively. $d)$ The ratio of the Hubble expansion rate within $\Lambda$MDM with neutrino-CC interaction and $\Lambda$CDM (solid) as well as $\Lambda$MDM and $\Lambda$CDM (dotted), respectively.}
\end{figure}

In the redshift range $z\sim 1400\div 3900$ a large fraction of neutrino energy-momentum gets transferred to the CC and thus the acceleration factor
~$q=\frac{\ddot{a}a}{\dot{a}^{2}}$ 
indicates a prolonging of the radiation dominated epoch ($q\approx-1$) with respect to standard cosmology. In other words, temporarily, the expansion is decelerating to a greater extend, as shown in Fig.~\ref{fig:QF-OmegaVz}$c)$. The reason for this background behavior is that the coupled negative vacuum energy and the decaying neutrinos act as a very stiff system as illustrated by the corresponding effective equation of state~$\omega_{{\rm eff}}$ in Fig.~\ref{fig:QF-OmegaVz}$b)$. In the final stages of the interaction the universe quickly approaches the standard evolution of a $\Lambda$~Mixed~DM model ($\Lambda$MDM) again, which corresponds to a $\Lambda$CDM model including neutrino masses. Therefore, the transition from the radiation dominated to the matter dominated epoch is much sharper than in the non-interacting case. As of then, the CC exhibits its standard value and redshifts with an equation of state of $\omega=-1$, while the neutrino energy density dilutes according to the standard law, however, its magnitude is reduced by a factor~$30$. Note that in the~$m_\nu=2.2\,{\rm eV}$ case this happens before recombination. 

Our third example shows neutrinos with degenerate masses~$m_\nu=0.5\,{\rm eV}$. In this case, the energy-momentum transfer starts after the time of recombination at approximately~$z=900$, and in comparison to the case of~$m_\nu=2.2\,{\rm eV}$ has less impact on the background evolution due to the smaller neutrino mass. Nevertheless, Fig.~\ref{fig:QF-OmegaVz} suggests that the integrated effects on the expansion rate from today to the time of recombination have an in principle measurable influence on cosmological observations, cf. Sec.~\ref{sec:Pheno}. 

The results discussed so far should also be applicable for other fields of higher masses as long as they are not interacting too strongly. Then, the mechanism starts at earlier times, when~$\Lambda$ was even more negative. However, as argued above, we expect the CC to be sub-dominant also during these stages and thus not to seriously interfere with standard Big Bang cosmology. Let us justify this statement by explicitly calculating the maximum of the ratio between the shift~$\Delta\Lambda$ of the running CC and the energy density~$\rho$ of the interacting fermions, 
\begin{equation}
 r:=\frac{\Delta\Lambda}{\rho} =\frac{3}{32}\left(\frac{m}{y_e(t)T}\right)^3\ln{\frac{y_e(t)T}{\mu_e}}.
\end{equation}
Note that the number of fermionic DOFs drops out. Since the maximum occurs close to the end of the interacting phase, we have employed the energy density and RG scale of a Fermi sphere from Eqs.~(\ref{eq:QF-rho-sph-FD}) and~(\ref{eq:QF-sph-mu-rho}). In addition, the final scale~$\mu_{e}^2=\frac{257}{256}m^2$ is fixed by~$y_eT_e=m/16$. Therefore, the maximum of~$r$ is acquired for~$y\,T/m\approx 0.11$, and its value~$r_{\rm max}\approx 0.57$ turns out to be independent of the field mass. Thus, let us apply this upper bound to the early universe. Accordingly, deep in the radiation dominated epoch, the total energy density~$\rho_{\rm tot}$ is dominated by many relativistic particle species such as photons, neutrinos, electrons and so on. In case one of these species becomes non-relativistic, it induces a vacuum shift which maximally corresponds to a small fraction of~$\rho_{\rm tot}$. Later on, however, when only few relativistic species are left over, this fraction can be as large as~$r_{\rm max}$. Therefore, the running vacuum energy shows a tracking behavior, which becomes better and better as time advances.

Despite the smallness of~$r$, one might still worry about early dark energy constraints at BBN. At this time, electrons and positrons should be subject to our mechanism and the induced changes in the CC could harm the background evolution. However, from particle species like e$^\pm$ that annihilate after their decoupling from the thermal bath, there is not much energy-momentum left over that can be transferred to the CC. Therefore, the interaction of the particles with the CC effectively ceases at this moment, and the induced vacuum shift is much smaller compared to a fully developed mechanism.

It is remarkable that even though we cannot explain the smallness of the current CC, we have found a mechanism that keeps its quantum corrections under control during the cosmological evolution. Additionally, in contrast to models of tracking quintessence, the considered scenario only requires a single free parameter,~$\Lambda_0$. It is also worth noting that the final vacuum shift induced by the neutrinos is of comparable magnitude as the present CC value, although both quantities are apparently not directly related to each other.

\section{Phenomenological implications\label{sec:Pheno}}
\subsection{Relaxing cosmological neutrino mass bounds}

In this section, we will consider the consequences arising from the non-standard neutrino interaction with the CC for the inference of neutrino mass bounds from current as well as from future cosmological measurements such as of the CMB and of LSS. 

We will start by extending the discussion of the last section on modifications to the background evolution caused by the interaction between the neutrinos and the CC. This will allow us to give a qualitative estimate of its impact on the spectrum of the CMB fluctuations and of the arising possible restrictions on the masses of interacting neutrinos. Furthermore, it will set the stage for our subsequent (semi-)analytical analysis in Sec.~\ref{sec:LSS} which demonstrates that within the considered scenario all current bounds on neutrino masses derived from LSS measurements can be evaded. A detailed numerical analysis can be found elsewhere~\cite{WIR}.

\subsubsection{Relaxing neutrino mass bounds from CMB data}

As already mentioned in the last section, the background evolution is modified in the presence of the considered non-standard neutrinos interaction due to the following two colluding effects. The first one results from the stiffness of the effective equation of state of the neutrino -- CC system during the interacting phase which tends to prolong the radiation dominated regime as argued before. The second effect on the expansion rate arises generally if the $\Lambda$CDM model is extended to include neutrino masses ($\Lambda$MDM)~\cite{Lesgourgues:2006nd}. However, as discussed in the following, its impact turns out to be much less important in the presence of the considered non-standard interaction. 

In general, while massive neutrinos are still relativistic at sufficiently early times and are thus counted as radiation, after the non-relativistic transition they contribute to the dark matter. This implies that the relative density of other species or the spatial curvature today have to get modified in comparison to a universe with massless neutrinos. However, for interacting neutrino the energy-momentum transfer to the CC becomes efficient as soon as they turn non-relativistic. This ensures that within a relatively short phase of interaction the neutrino energy density is reduced to $1/30$ of its standard value, cf.\ Eq.~(\ref{eq:QF-mapp}). Accordingly, in contrast to the standard case, the contribution of interacting neutrinos to the total energy density today is $\Omega_{\nu}h^2<2.4\times 10^{-3}$ (for $\sum m_{\nu}<6.6$ eV as required by tritium bounds~\cite{Weinheimer:2003fj,Lobashev:2001uu}). Hence, for a flat universe, this demands a negligible reduction of the dark matter density compared to the massless neutrino case. As a consequence, the usual effect of neutrino masses on cosmology of postponing the matter-radiation equality is much less pronounced for interacting neutrinos. However, the interaction with the CC has an influence on the background evolution and thus on the time of transition from radiation to matter domination, where its impact grows with increasing neutrino mass. This can be seen by comparing the evolution of the acceleration parameter $q$ with and without the interaction in Fig.~\ref{fig:QF-OmegaVz}. While the radiation dominated regime is prolonged, the transition to the matter-dominated regime happens faster in the presence of the interaction, since it quickly causes neutrinos to become highly non-relativistic. This is also demonstrated by Fig.~\ref{fig:QF-OmegaVz}, where the ratio of the Hubble expansion rate within $\Lambda$MDM with neutrino-CC interaction and $\Lambda$CDM with massless neutrinos is compared to the corresponding ratio within $\Lambda$MDM without the new interaction and $\Lambda$CDM. Especially the effects on the integrated expansion rate are expected to be smaller for interacting neutrinos as suggested by Fig.~\ref{fig:QF-OmegaVz}. For the CMB power spectrum, however, it is mainly the integrated expansion rate, which determines the important scales that set the location and the height of the acoustic peaks~\cite{Lesgourgues:2006nd}. More precisely, the sound horizon at recombination is given by, 
\begin{equation}
r_s=\int\limits_{z_{\rm{rec}}}^{\infty} \frac{c_s\,\rm{d}z}{H(z)}
\end{equation}
with $c_s$ denoting the sound speed of the baryon-photon fluid. The observable angular scale of the acoustic oscillations depends on an integral from the present to recombination involving the density and the equation of state of the running CC~\cite{Melchiorri:2002ux,Hannestad:2004cb}. While keeping all other parameters fixed, with respect to a $\Lambda$CDM model, in principle a linear shift in the angular position of the CMB features could be induced. However, the effective $\omega_{\rm eff}$ only differs from $\omega=-1$ during the relatively short phase of interaction to an extent which decreases with the neutrino mass. Furthermore, already much before the dark energy dominated regime the energy density in the running CC has essentially reached its present value (cf.~Fig.~\ref{fig:QF-LambdaVz}). Accordingly, the integrated result generally will not differ much from the $\Lambda$CDM case.\footnote{From similar arguments follows that the influence of the interaction on the magnitude of the integrated Sachs-Wolfe effect at low multipoles is expected to be small.} We note that secondary effects on the CMB acoustic peaks resulting from the free streaming of relativistic neutrinos (i.e.\ a phase shift and an amplitude reduction at large multipoles)~\cite{Bashinsky:2003tk} are expected to essentially be unaltered in the considered scenario, since in the relativistic neutrino regime the coupling between neutrinos and the CC is very strongly suppressed (cf.~Sec.~\ref{sec:RGE}). 

Hence, in summary, qualitatively, compared to the standard $\Lambda$CDM case with massless neutrinos, the modifications to the CMB spectrum arising from the inclusion of neutrino masses are less pronounced in the presence of the non-standard neutrino interaction. Therefore, for interacting neutrinos the current neutrino mass bound of $\sum m_{\nu}=2\div 3$ eV (at $95\%$ c.l.) gained from WMAP3 data alone is expected to be relaxed (cf.~\cite{Lesgourgues:2006nd} and references therein, see~\cite{Fukugita:2006rm} for the tightest bound). Thus, a priori, the considered scenario does not seem to be excluded by current CMB data, even if the sum of neutrino masses is in the super-eV range as allowed by current tritium experiments. A detailed analysis of the resulting modifications to the CMB spectrum and the arising bounds on the masses of interacting neutrinos goes beyond the scope of this paper and will be investigated elsewhere~\cite{WIR}.

In the next subsection we will discuss the consequences of the non-standard neutrino interaction on the inference of neutrino mass bounds from LSS measurements. 

\subsubsection{Evading current neutrino mass bounds from LSS measurements\label{sec:LSS}}

Structure formation is sensitive to neutrino masses through kinematic effects caused by the neutrino free streaming as characteristic for hot and warm dark matter~\cite{Hu:1997vi,Hu:1997mj}. The associated important length scale corresponds to the typical distance neutrinos can propagate in time $t$ in the background spacetime and is governed by their mean velocity $\langle v\rangle$, 
%\be\label{FS}
%\lambda_{\rm FS}=a(t)\int\limits_{0}^{t}\frac{\langle v\rangle}{a(t')} dt'=2\pi\frac{a(t)}{k_{\rm FS}}=2\pi\sqrt\frac{2}{3}\frac{\langle v\rangle}{H(t)},
%\ee
\be\label{FS}
\lambda_{\rm FS}(t)=2\pi\frac{a(t)}{k_{\rm FS}}=2\pi\sqrt\frac{2}{3}\frac{\langle v\rangle}{H(t)},
\ee
where $\lambda_{\rm FS}/a$ and $k_{\rm FS}$ respectively denote the comoving free streaming length and wavenumber of the neutrinos. As a consequence, on the level of perturbations neutrinos do not aid to the gravitational clustering of matter on scales below the horizon when they turn non-relativistic. However, their energy density $\rho_{\nu}$ contributes to the homogenous background expansion through the Friedmann equation. Through the metric source term in the perturbed Einstein equation, this imbalance leads to a slow down of the growth of matter perturbations on small scales (cf. Ref.~\cite{Lesgourgues:2006nd} for a recent review). It crucially depends on the relative fraction provided by neutrinos to the total energy density $\rho_{\rm m}$ of matter\footnote{Note that the formula in Eq.~(\ref{f}) approximately also holds true in case a very light neutrino state has not turned non-relativistic yet, because in this case its relative contribution to the energy density is negligible anyway~\cite{Lesgourgues:2006nd}.},
\be\label{f}
{\mathfrak{f}}_{\nu}\equiv\frac{\rho_{\nu}}{\rho_{\rm m}}\approx \frac{\sum m_{\nu}}{15\,{\rm eV}},
\ee
where $\rho_{m}$ comprises the energy densities of cold dark matter, baryons and neutrinos. 

Assuming ${\mathfrak{f}}_{\nu}\ll 1$, on scales smaller than the horizon when neutrinos turn non-relativistic, the resulting net effect on the present day matter power spectrum $P_m(k)$ for a normalization at $k\longrightarrow 0$ is a suppression of~\cite{Hu:1997mj}
\be\label{P}
\frac{\Delta P_m(k)}{P_m(k)}\simeq -8{\mathfrak{f}}_{\nu}.
\ee 
 
In the following we will demonstrate that the proposed interaction between relic neutrinos and the CC allows to completely evade present neutrino mass limits deduced from LSS data below the tritium bound $m_{\nu}<2.2$ eV~\cite{Weinheimer:2003fj,Lobashev:2001uu}, because the neutrino energy density gets reduced below the current sensitivity of LSS measurements on all accessible scales. However, at the end of the section we will argue that both for the Planck mission as well as for weak lensing or high redshift galaxy surveys it seems feasible to probe the proposed non-standard neutrino interaction. To this end, we will investigate the modifications to the characteristic scale for free streaming signatures induced by the considered neutrino interaction. 

Let us start by noting that for structure formation the interaction generally has a negligible influence on the expansion rate independent of the actual neutrino mass scale realized in nature. Namely, if the interaction takes place at late times during structure formation, the corresponding neutrino masses and the amount of energy-momentum transferred to the CC are too small to lead to an observable effect on $H$. However, in case the interaction has an impact on the expansion rate for sufficiently large neutrino masses, then it ceases already much before structure formation. Accordingly, in the following analysis we will adopt $\tilde{H}(z)\sim H(z)$, where a tilde throughout this work labels quantities which assume an interaction between neutrinos and the CC. 

\begin{figure}
\vspace* {0.0in}
\begin{center}
\includegraphics*[bbllx=26pt,bblly=238pt,bburx=584pt,bbury=701pt,height=8.3cm]{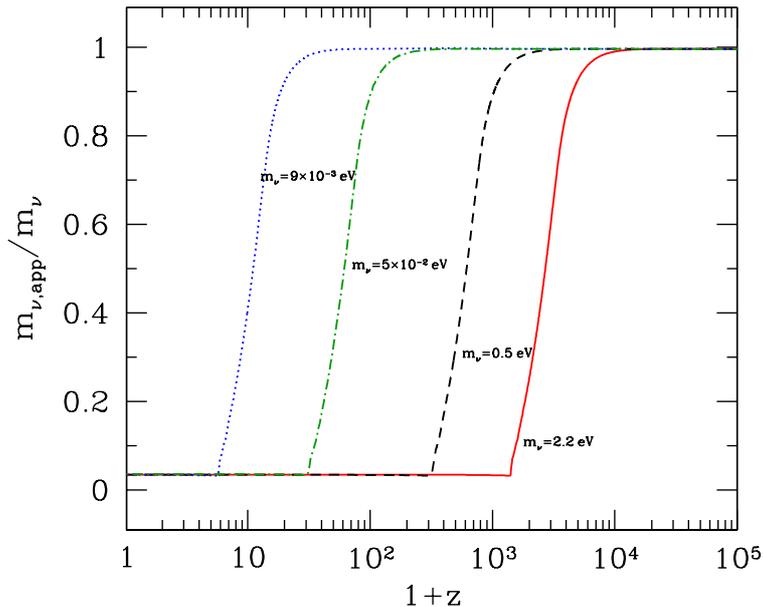}
\caption[]{\label{fig:mapp}The apparent neutrino mass as a function of redshift as given by Eq.~(\ref{eq:mapp}) for $m_{\nu}=2.2$ eV, $m_{\nu}=0.5$ eV, $m_{\nu}=5\times 10^{-2}$ eV and $m_{\nu}=9\times 10^{-3}$ eV. Note that the ratio $m_{\rm app}/m_{\nu}\simeq 1/30$ is independent of the neutrino mass for $1+z<m/(9T_0)$.}
\end{center}
\end{figure}

Let us now proceed by comparing the fractions ${f}_{\nu}$ and $\tilde{f}_{\nu}$ provided by neutrinos to the total matter density with and without non-standard interaction in the non-relativistic neutrino regime, $T_{\nu}<m_{\nu}/3$, where $\rho_{\nu}(z)\simeq m_{\nu}n_{\nu}(z)$. In accordance with Eq.~(\ref{eq:QF-mapp}), we define the apparent mass of the neutrinos also for times before the interaction ends,
\be\label{eq:mapp}
m_{\rm app}(T_{\nu})\equiv m_{\nu}\frac{\tilde{\rho}_{\nu}(T_{\nu})}{\rho_{\nu}(T_{\nu})}=m_{\nu}\frac{\tilde{n}_{\nu}(T_{\nu})}{n_{\nu}(T_{\nu})}.
\ee 
Here, $m_{\rm app}$ represents the mass of a non-relativistic neutrino an observer would infer from the extent of a neutrino-induced suppression of the matter power spectrum, if he assumes the standard scaling of $\rho_{\nu}$ for a Fermi-Dirac distribution. Per definition $m_{\rm app}$ absorbs the additional non-standard redshift dependence common to $\tilde{n}_{\nu}$ and $\tilde{\rho}_{\nu}$ caused by the interaction with the CC. We have plotted $m_{\rm app}$ as a function of the redshift for different possible values of the neutrino mass in Fig.~\ref{fig:mapp}, which demonstrates that $m_{\rm app}$ becomes constant for $T_{\nu}\lwig m_{\nu}/9$ when the energy exchange with the CC has ceased. Correspondingly, after neutrinos have effectively decoupled from the CC,
their energy density again obeys the standard scaling law with redshift, $\tilde{\rho}_{\nu}\propto (1+z)^3$. However, compared to the energy density of a non-interacting neutrino of mass $m_\nu$, its absolute magnitude is reduced by a (mass independent) factor of $m_{\nu}/m_{\rm app}\simeq 1/30$ (cf. Eq.~(\ref{eq:mapp}) and Fig.~\ref{fig:mapp}). Consequently, since tritium $\beta$ decay experiments constrain neutrino masses to be smaller than $2.2$ eV~\cite{Weinheimer:2003fj,Lobashev:2001uu}, the apparent neutrino mass has to be smaller than $m_{\rm app}<2.2/30\,{\rm eV}\simeq 7\times 10^{-2}$ eV for $T_{\nu}<m_{\nu}/9$. Thus, according to Eq.~(\ref{f}), the suppression of the matter power spectrum on scales below the neutrino free streaming becomes proportional to,
\be
\tilde{\mathfrak{f}}_{\nu}=\frac{m_{\rm app}}{m_{\nu}}\mathfrak{f}_{\nu}\approx\frac{1}{30}\mathfrak{f}_{\nu} \approx\frac{\sum m_{\nu}}{450\,{\rm eV}}<0.015.
\ee
This result demonstrates that for $\sum m_{\nu}<6.6$ eV the interaction with the CC reduces the efficiency of the suppression of the power spectrum achieved by neutrino free streaming below the current sensitivity of LSS measurements. It should be noted that this result is independent of the actual scales where neutrino free streaming is relevant. It thus implies that within the considered scenario all current bounds on $m_{\nu}$ from LSS data can be evaded.

However, in the next subsection, we will discuss the characteristic free streaming signatures of interacting neutrinos and see how they can possibly be revealed by future cosmological probes.

\subsection{Characteristic free streaming signatures}

Future weak lensing and also high redshift galaxy surveys in combination with Planck data promise a considerable increase in the sensitivity to neutrino mass to $\sigma(\sum m_{\nu})\lwig 0.05$ eV~\cite{Hannestad:2006as,Hannestad:2007cp}. For instance, surveys of weak gravitational lensing of distant galaxies directly probe the matter distribution without having to rely on assumptions about the luminous versus dark matter bias in contrast to conventional galaxy redshift surveys. The improvement of the sensitivity is largely due to the possibility of breaking parameter degeneracies in the matter power spectrum in particular by the help of tomographic information on the power spectrum resulting from a binning of the source galaxies by redshift. Accordingly, while existing LSS surveys are mainly sensitive to the transition region in the power spectrum, where free streaming effects start to become important, these future probes at different redshifts can accurately measure a broader interval of wavenumbers extending into the present non-linear regime (e.g.\ at the peak of the sensitivity of weak lensing surveys $k\sim 1\div 10h\,\rm {Mpc}^{-1}$ at $z=0.5$~\cite{Hannestad:2006as}, while galaxy surveys probe scales $4.5\times 10^{-3}h\,\rm {Mpc}^{-1}\div 1.5 h\,\rm {Mpc}^{-1}$ at $0.5<z<6$~\cite{Hannestad:2007cp}). By these means, they are in principle sensitive to the characteristic step-like effects on the power spectrum arising from neutrino free streaming and as a result to much smaller ${\mathfrak{f}}_{\nu}$. As we will see in the following, this offers the possibility to test the considered scenario by these future probes. The reason is that the interaction between neutrinos and the CC turns out to not only reduce the energy density, but also the free streaming scale of the neutrinos for a given mass. Accordingly, the free streaming signatures tend to be shifted towards the non-linear regime as we will demonstrate in the following.  

To this end, let us proceed by comparing the characteristic velocities of neutrinos which determine the free streaming scale according to Eq.~(\ref{FS}) with and without non-standard neutrino interaction. Starting with the standard case, the neutrinos are assumed to be Fermi-Dirac distributed, the average neutrino velocity corresponds to the thermal velocity $\langle v\rangle$,
\be
\langle v\rangle \equiv \frac{\langle p\rangle}{m_{\nu}}\simeq \frac{3T_{\nu}}{m_{\nu}},
\ee
where $\langle p\rangle$ denotes the mean momentum of freely propagating relic neutrinos. However, as described in the last section, the presence of an interaction with the CC tends to remove high momenta from the neutrino spectrum. As a consequence, in this case the average neutrino velocity gets reduced and the transition to the non-relativistic regime is speeded up compared to non-interacting neutrinos. Since a detectable suppression of the power spectrum by future surveys corresponding to $\sum m_{\rm app}\gwig 0.03$ eV for the most ambitious projects~\cite{Song:2004tg,Lesgourgues:2006nd} requires the sum of neutrino masses in the interacting case still to be in the super-eV range, we can safely assume the interaction with the CC has ceased already much before structure formation. Thus, after the final stage of the interaction, the mean neutrino velocity $\langle \tilde{v}\rangle$ of the maximally degenerated neutrino distribution with Fermi-momentum $p_{\rm F}= y_e T_{\nu}=4/3\langle \tilde{p}\rangle$ can be expressed as,
\be
\langle \tilde{v}\rangle \equiv \frac{\langle \tilde{p}\rangle}{m_{\nu}}\simeq \frac{3}{4}\frac{y_e T_{\nu}}{m_{\nu}}.
\ee
Accordingly, by the help of Eq.~(\ref{FS}) and Eq.~(\ref{ye}) we arrive at the relation,
\be\label{tFS}
\tilde{k}_{\rm FS}=k_{\rm FS}\frac{\langle v\rangle}{\langle\tilde{ v}\rangle}\simeq\frac{64}{9}\,k_{\rm FS}\simeq 7.11\,k_{\rm FS},
\ee
where $\tilde{H}(z)\simeq H(z)$ was assumed as justified above. Consequently, since on average the interacting neutrinos are slower, their comoving free streaming wavenumber $\tilde{k}_{\rm FS}$ according to Eq.~(\ref{tFS}) is roughly an order of magnitude larger than in the standard case. We would like to stress that this result holds independently of the actual neutrino mass scale realized in nature for redshifts $1+z<m_{\nu}/(9T_{0})$.

For neutrinos becoming non-relativistic during matter domination, the comoving free streaming wave\-number passes through a minimum at the time of the non-relativistic transition. Thus, in the standard case, the matter power spectrum is suppressed on all scales much larger than~\cite{Lesgourgues:2006nd},
\be\label{kFS}
k_{\rm nr}\simeq 1.8\times 10^{-2}\,\Omega^{1/2}_{\rm m}\left(\frac{m_{\nu}}{1\,{\rm eV}}\right)^{1/2} h {\rm Mpc}^{-1}.
\ee 
In contrast, the free streaming signatures of interacting neutrinos are expected to influence all wavenumbers larger than the minimal comoving wavenumber at $T_{\nu}\simeq m_{\nu}/9$,
\be\label{tkFS}
\tilde{k}_{\rm nr}\simeq 0.23\, \Omega^{1/2}_{\rm m}\left(\frac{m_{\nu}}{1\,{\rm eV}}\right)^{1/2}h {\rm Mpc}^{-1}.
\ee
For comparison, for $\Omega_{\rm m}\simeq 0.3$ and $0.046<m_{\nu}<2.2$ eV this corresponds to $2.1\times 10^{-3}h\,{\rm Mpc}^{-1}<k_{\rm nr} <1.5\times 10^{-2}h\,{\rm Mpc}^{-1}$ in contrast to $2.7\times 10^{-2}h\,{\rm Mpc}^{-1}<\tilde{k}_{\rm nr} <0.18 h\,{\rm Mpc}^{-1}$. Clearly, the free streaming signatures of interacting neutrino are shifted towards the non-linear regime. 

Hence, in summary, a clear signature for the considered scenario is provided by the non-standard correspondence between the extent of the suppression of the matter power spectrum due to neutrino free streaming and the corresponding scales which are affected. In other words interacting neutrinos appear lighter, however, at the same time can free stream a shorter distance in a Hubble time than in the standard case. 

Before turning to the prospects of testing the considered scenario by future probes, let us mention another characteristic signature expected to arise in its framework if $\sum m_{\nu}$ is in the super-eV range. As argued above, in this case the new neutrino interaction has an impact on the time of matter-radiation transition whose extent grows with increasing neutrino mass. Accordingly, this leads to a translation of the turning point in the matter power spectrum in comparison to the $\Lambda$CDM case, because the scales entering the horizon at equality have a different size. However, the evolution of the acceleration parameter $q$ is characteristically altered compared to the case of non-interacting massive neutrino as described above. Thus, it is expected to lead to a shift of the turning point compared to both the $\Lambda$CDM and the $\Lambda$MDM case whose location, however, has to be determined numerically~\cite{WIR}.

\section{Probing the scenario\label{sec:Probing}}

\subsection{Future cosmological probes and interacting neutrinos}

In this subsection, we will give an outlook how combined future weak lensing tomography surveys and CMB Planck data could be used to probe non-standard neutrino interactions and in particular the proposed new interaction with the CC. As it turns out the achievable explanatory power about the underlying interaction increases with the neutrino mass.
  
In general, the combined projected sensitivity of lensing surveys and future CMB data is predicted to reach $\sigma(\sum m_{\nu})\gwig 0.03$ eV for non-interacting neutrinos depending on the employed data sets and on the number of free parameters of the model (see Ref.~\cite{Lesgourgues:2006nd} and references therein). According to solar and atmospheric neutrino oscillation experiments the neutrino mass squared differences are given by $\Delta m^2_{\rm atm}=|\Delta m^2_{23}|\simeq 2.4\times 10^{-3}\,{\rm eV}^2$ and $\Delta m^2_{\rm sun}=\Delta m^2_{12}\simeq 7.9\times 10^{-5}\,{\rm eV}^2$ (e.g.\ Ref.~\cite{Strumia:2006db}). Correspondingly, even if the lightest neutrino is massless, the most ambitious projects provide a $2\sigma$ sensitivity to the minimal normal hierarchy with $\sum m_{\nu}=0.06$ eV and could thus distinguish it from the minimal inverted hierarchy corresponding to $\sum m_{\nu}=0.1$ eV. Accordingly, non-standard neutrino interactions could be revealed by these future probes, if neutrino free streaming signatures are not observed (see e.g.~\cite{Beacom:2004yd}). Interpreted in terms of the considered scenario, this would indicate the sum of neutrino masses to be in the sub-eV range such that the corresponding non-standard relic neutrino density is reduced below the sensitivity of these experiments (cf. Sec.~\ref{sec:Pauli-Blocking} for an alternative, complementary way of testing the nature of the underlying interaction in the laboratory which is independent of the actual neutrino mass scale, since possible modifications to the neutrino distribution function are probed.)

However, if $\sum m_{\nu}=1.2\div 2.7$ eV as suggested by some members of the Heidelberg-Moscow collaboration~\cite{KlapdorKleingrothaus:2004wj}, then it seems feasible for future cosmological probes to unambiguously reveal the proposed new neutrino interaction with the CC and to verify the Heidelberg-Moscow claim within this scenario as discussed in the following. 

Since CMB data and LSS surveys give us a snap shot of the universe at different times, they offer the possibility to compare the influence of neutrinos on cosmology before (or while) and after the interchange with the CC, respectively.

If the sum of neutrino masses is in the super-eV range, the phase of interaction between the neutrinos and the CC overlaps with the epoch of recombination and thus during this time leads to a characteristic modification of the expansion rate as described above. Within a $7$ parameter $\Lambda$MDM framework, CMB data is able to provide a bound on neutrino masses which is independent of LSS data due to the absence of a parameter degeneracy~\cite{Lesgourgues:2006nd}. Considering that the projected sensitivity of Planck data alone is $\sigma(\sum m_{\nu})\lwig 0.48$ eV for non-interacting neutrinos (see Ref.~\cite{Lesgourgues:2006nd} for a review), it thus seems feasible to verify the Heidelberg-Moscow claim of $\sum m_{\nu}>1.2$ eV for interacting neutrinos by CMB data alone. 

In addition, this result could be tested in a complementary way by future weak lensing or high redshift galaxy surveys which probe the relic neutrino background after the energy-momentum exchange with the CC has considerably reduced its energy density. This, however, implies $\omega=-1=\rm{const.}$ for the CC during structure formation and thus the absence of an apparent degeneracy between $m_{\nu}$ and $\omega$ usually arising from a possible time-variation of $\omega$. Thus, it seems reasonable to compare to the projected sensitivity of future LSS probes within a $7$ parameter $\Lambda$MDM framework. Since $\sum m_{\nu}>1.2$ eV would correspond to $\sum m_{\rm app}>0.04$ eV, the most ambitious projects (see Refs.~\cite{Song:2004tg,Lesgourgues:2006nd}) promise to be able to verify the whole neutrino mass range indicated by part of the Heidelberg-Moscow collaboration for neutrinos interacting with the CC. In order to be proven right, the extent of the suppression of the power spectrum would require $\sum m_{\rm app}>0.04$, but the corresponding scales of the power spectrum affected by neutrino free streaming would be $>\tilde{k}_{\rm nr}=8\times 10^{-2}h{\rm Mpc}^{-1}$ according to Eq.~(\ref{tkFS}) instead of $>2\times 10^{-3}h{\rm Mpc}^{-1}$ as expected from the standard relation in Eq.~(\ref{kFS}). In summary, if the sum of neutrino masses is in the super-eV range, the prospects seem good to probe the considered scenario by future LSS surveys combined with future CMB data. In addition, in this case the absolute neutrino mass scale could be ascertained by identifying the predicted characteristic non-standard neutrino free streaming signatures on the power spectrum.

\subsection{Pauli-blocking effects and interacting neutrinos\label{sec:Pauli-Blocking}}

Before we conclude this section, we would like to mention another interesting way of testing the considered scenario in the laboratory which is complementary to cosmological measurements. Importantly, it turns out to be independent of the actual neutrino mass scale realized in nature. Namely, proposed experiments searching for Pauli-blocking effects~\cite{Cocco:2007za,Takahashi:2007ec} resulting from the presence of the \CnuB are in principle sensitive to possible modifications of the neutrino distribution function~$f$ induced by the neutrino interaction with the CC as discussed in the following. For instance, in the experiment proposed in Ref.~\cite{Takahashi:2007ec}, the pair production of neutrinos with very low momenta~$p\approx 0$ is investigated, where the event rate is sensitive to the Pauli blocking factor~$(1-f)$. In this context, a standard FD distribution~$f_{\rm{FD}}$ for the \CnuB would lead to a suppression factor~$(1-f_{\rm{FD}})\simeq 1/2$ in the event rate. In contrast, in our set-up, the
distribution of the relic neutrinos is maximally degenerate with all states occupied up to the Fermi momentum $p_{\rm F}$ after the interaction with the CC has ceased, thus leading to a sharp-edged step function~$(1-f(p))=\Theta (p-p_{\rm F})$ for the Pauli blocking term. Consequently, neutrinos cannot be produced with momenta below the current Fermi momentum~$p_{\rm F}=p_e T_0/T_e=y_e T_0$, while the production of neutrinos with larger momenta is not suppressed at all. Compared to the FD case this implies a much sharper pair production threshold and in addition a shift in the threshold energy due to the non-vanishing Fermi momentum~$p_{\rm F}\approx 1.0\times 10^{-4}$~eV.

\section{Conclusions and Outlook}

The claimed evidence for neutrinoless double beta decay translates into a neutrino mass bound of $\sum m_{\nu}>1.2$ eV at $95\%$~\cite{KlapdorKleingrothaus:2004wj} which is in tension with current neutrino mass bounds derived from cosmology~\cite{Goobar:2006xz,Seljak:2006bg,Feng:2006zj,Cirelli:2006kt,Hannestad:2006mi,Fogli:2006yq,Spergel:2006hy,Tegmark:2006az,Zunckel:2006mt,Kristiansen:2006ky}. In this work, we have shown that cosmological neutrino mass bounds can be relaxed and brought into agreement with the Heidelberg-Moscow claim, if a newly proposed neutrino interaction is taken into account which can be tested by future CMB and LSS measurements. It acts between relic neutrinos and the CC and arises from the zero-point energy contributions of the neutrino quantum fields. Since they induce a scale-dependence of the vacuum energy via renormalization group effects, the CC becomes time-dependent as long as the renormalization scale runs. In this case, an energy-momentum exchange between the relic neutrinos and the CC is implied through the Bianchi identity.

We have studied in detail the time-evolution of the coupled system, in particular its consequences for the dynamics of the CC, the spectrum and the abundance of relic neutrinos and its impact on the interpretation of cosmological measurements. We have found that owing to the relative smallness of neutrino masses, the interaction becomes of dynamical influence at late times when neutrinos turn non-relativistic, $T_{\nu} \simeq m_{\nu}/3$. Moreover, it effectively ceases after a relatively short period, at a temperature also set by the neutrino mass, $T_{\nu} \simeq m_{\nu}/9$. This decoupling behavior can be attributed to our particle specific choice for the renormalization scale. Since it was taken to be the average available neutrino energy which asymptotically runs to a value slightly above the neutrino mass, its evolution and thus the running of the CC effectively cease in the non-relativistic neutrino regime. By taking into account other fermions of higher masses, it was also shown that the corresponding vacuum contributions are becoming more and more important at late times, whereas they are completely sub-dominant in the early universe. During the cosmological evolution, the CC is therefore approaching a tracking regime at late times.

Accordingly, in the relativistic neutrino regime after decoupling we have found the neutrino distribution to essentially maintain its standard Fermi-Dirac form, while the CC runs only logarithmically with time, but with a smaller value than today. However, when turning non-relativistic, the neutrinos were found to efficiently transfer energy-momentum to the CC, thus driving its energy density to its value as measured today. Since as a result, high neutrino momenta turned out to be removed from the neutrino spectrum while low momenta got enhanced, remarkably, the interaction was found to deform the neutrino momentum distribution into a maximally degenerate form. In comparison to Fermi-Dirac distributed neutrinos of equal mass, the mean neutrino momentum decreased by one order of magnitude implying a reduction of the neutrino energy density to $1/30$ of its standard value after the interaction has ceased. Interpreted in terms of the standard relation between the neutrino energy density and mass, the non-relativistic neutrinos redshift as CDM, but appear lighter $m_{\rm app}=1/30\,m_{\nu}$. Accordingly, the presence of the non-standard neutrino interaction was found to considerably relax cosmological neutrino mass bounds. In particular, within the considered scenario current galaxy redshift surveys were shown not to be sensitive at all to neutrino masses below the upper bound $m_{\nu}=2.2$ eV from tritium experiments~\cite{Weinheimer:2003fj,Lobashev:2001uu} and could thus be evaded. In addition, we have argued that the present comparatively mild neutrino mass bound from WMAP3 year data of $\sum 2\div 3$ eV at $2\sigma$ can be relaxed in the presence of the proposed neutrino interaction with the CC. The reason primarily is that in comparison to standard $\Lambda$CDM cosmology, the characteristic integrated effects on the expansion rate turned out to be smaller than in the case of a $\Lambda$MDM model. 

However, we have proposed possible tests for the non-standard neutrino interactions with the CC by future cosmological probes as well as by laboratory based experiments. In the former case, the explanatory power on possible new physics turned out to increase with the neutrino mass. Since future weak lensing surveys probe a broad range of comoving wavenumbers at different redshifts, they provide unbiased, tomographic information about the matter power spectrum. Combined with Planck data the most ambitious projects promise a $2\sigma$ sensitivity to the minimal value in case of a normal hierarchy ($\sum m_{\nu}\sim0.06$ eV)~\cite{Lesgourgues:2006nd} such that a non-observation of neutrino free streaming signatures would provide a hint for non-standard neutrino interactions. Interpreted in terms of the considered scenario this would mean that the interaction has reduced the neutrino energy density below the sensitivity of these future probes implying the sum of neutrino masses to be in the sub-eV range. 

In this case it might still be feasible for proposed laboratory based experiments to reveal the unique signatures of a possible interaction with the CC by directly probing the neutrino distribution function through Pauli-blocking effects~\cite{Cocco:2007za,Takahashi:2007ec}. 

If, on the other hand, the sum of neutrino masses is larger than $\sum m_{\nu}>1.2$ eV as suggested by part of the Heidelberg-Moscow collaboration, we found future LSS probes also to be able to unambiguously reveal the characteristic free streaming signatures of neutrinos having interacted with the CC. Namely, Fermi-Dirac distributed neutrinos can free stream larger distances the lighter they are. However, as a unique signature of the energy-exchange with the CC, neutrinos become slower than in the absence of the interaction and thus free stream a shorter distance, while at the same time they appear lighter due to their reduced abundance. 

In addition, it seems likely that future Planck data alone could provide a complementary test for the Heidelberg-Moscow result within the considered scenario, since in this case the interacting phase overlaps with the recombination epoch. Considering that the projected sensitivity in the absence of the neutrino-interaction is $\sigma(\sum m_{\nu})\lwig 0.48$ eV (see e.g. Ref.~\cite{Lesgourgues:2006nd} and references therein), it thus seems feasible by Planck data to trace the characteristic modifications to the expansion rate resulting from the neutrino -- CC interaction. 

Let us in the following give an outlook on possible extensions of the considered scenario. Since theories beyond the standard model predict other relics as warm or cold dark matter candidates, e.g.\ axions, gravitinos or supersymmetric partners of standard model fermions, their effects on the running of the CC relevant at energy scales of the order of their masses, could consistently be studied within the proposed renormalization scheme. Such an extension of the considered scenario is especially interesting, because the masses of bosonic species enter the RGE for the CC with an opposite sign as fermions. This might indicate that while the interaction with the CC generically causes the fermionic species to lose energy-momentum, in turn, the bosonic species would gain energy-momentum. Thus, presumably this would tend to increase the momenta of the bosonic distribution, making the bosonic dark matter candidate hotter, while the energy density in the CC would decrease. However, in order to be consistent with observations, within this scenario strong restriction on the particle masses are expected to arise or their existence could turn out to be excluded after all. 

After the posting of the first version of this paper, Ref.~\cite{Foot:2007wn} appeared. The corresponding authors claim that in the work on the RGE running of the CC~\cite{Shapiro:1999zt,Shapiro:2000dz} part of the one-loop corrections to the CC have been missed out, namely the explicit $\mu$ dependent part of the full one-loop effective action. It is furthermore argued that on including the missing part the CC does not formally run with the renormalisation scale $\mu$. In our opinion this claim is misleading, since it is precisely this allegedly missed out one-loop correction which gives rise to the non-zero beta-function for the CC in Eq.~(\ref{eq:QF-RGE}) (determined by the {\it partial} $\mu-$derivative). Moreover, in the framework of an RG improvement within the effective field theory approach taken in our work, it is as usual the (non-zero) beta-function which describes the RG evolution. Therefore, we do not see the appropriateness of our RG analysis of the CC invalidated by the above paper.

%%%%%%%%%%%%%%%%%%%%%%%%%%%%%%%%%%%%%%%%%%%%%%%%%%%%%%%%%%%%%%%%%%%%%%
\section*{Acknowledgments} %%%%%%%%%%%%%%%%%%%%%%%%%%%%%%%%%%%%%%%%%%%
%%%%%%%%%%%%%%%%%%%%%%%%%%%%%%%%%%%%%%%%%%%%%%%%%%%%%%%%%%%%%%%%%%%%%%

LS thanks Andreas Ringwald for advice, continuous support and
discussions and Alessandro Melchiorri and Gianpiero Mangano for 
helpful discussions. 

\section*{References}

\bibliographystyle{utphys}
\bibliography{Lnu-RGE-Biblio}

\end{document}